\documentclass[10pt,times,twocolumn]{article}
\usepackage{times,fullpage}
\usepackage{epsfig,url,amsmath,subfigure}

\begin{document}

\title{Dynamic Localization Protocols for Mobile Sensor Networks} 

\author{Sameer Tilak, Vinay Kolar, Nael B. Abu-Ghazaleh and Kyoung-Don Kang \\
        Dept. of CS, Binghamton University \\
        Binghamton, NY~~13902--6000 \\
    \url{{sameer,vinkolar,nael,kang}@cs.binghamton.edu}}

\maketitle

\begin{abstract}
The ability of a sensor node to determine its physical location within
a network (Localization) is of fundamental importance in sensor
networks. Interpretating data from sensors will not be possible unless
the context of the data is known; this is most often accomplished by
tracking its physical location.  Existing research has focused on
localization in static sensor networks where localization is a
one-time (or low frequency) activity.  In contrast, this paper
considers localization for mobile sensors: when sensors are mobile,
localization must be invoked periodically to enable the sensors to
track their location.  The higher the frequency of localization, the
lower the error introduced because of mobility.  However, localization
is a costly operation since it involves both communication and
computation. In this paper, we propose and investigate adaptive and
predictive protocols that control the frequency of localization based
on sensor mobility behavior to reduce the energy requirements for
localization while bounding the localization error.  We show that such
protocols can significantly reduce the localization energy without
sacrificing accuracy (in fact, improving accuracy for most
situations).  Using simulation and analysis we explore the tradeoff
between energy efficiency and localization error due to mobility for
several protocols.
\end{abstract}

\section{Introduction}
Localization is the ability of a sensor to find out its physical
 coordinates; this is a fundamental ability for embedded networks
 because interpretating the data collected from the network will not
 be possible unless the physical context of the reporting sensors is
 known.  In addition, localization is of importance in Moble Ad hoc
 NETworks (MANETs): several protocols utilize geographical information
 to improve operation (e.g.,~\cite{lar}).  Existing research has
 focused on addresing localization problem static sensor networks
 (sensors once deployed are stationary throughout
 life-time).

Localization may be carried out in one of several ways.  If the node
is equipped with a Global Positioning System (GPS) card, it can
determine its coordinated by receiving signals from a number of
sattelites.  Differential GPS requires that the node also receives
signals from nearby ground reference stations.  GPS cards
are often too expensive and/or power hungry for embedded micro-sensors
or even low end mobile devices such as PDAs.  In addition, GPS does
not work inside buildings where the Sattelite signals cannot be
received. Alternative localization approaches have been proposed to
allow nodes to learn their location either from neighboring nodes or
from reference beacons~\cite{Bulusu00a,priyantha00cricket}.  In these
approaches, the node has to communicate to/from beacons and/or
neighboring nodes.  For example, in one approach a node requiring
localization may broadcast a query to all beacons in range and then
receive replies from each of them allowing it to compute its location
as the center of gravity of the beacon locations.  Since these
approaches require communication, localization requires significant
energy.  

In this paper, we consider energy-efficient dynamic localization
 protocols for mobile wireless sensor devices.  More specifically, we
 are concerned with the problem of deciding when to invoke
 localization, regardless of the underlying localization mechanism.
 Since there is an energy cost involved in localization, we would like
 to minimize the localization frequency.  However, since the sensors
 are mobile, localization must be carried out with a frequency
 sufficient to capture the sensor location with acceptable error
 tolerance.  Although we focus primarily on sensors, the proposed
 algorithms also apply to other mobile node localization problems
 including MANETs and last hop network localization.

Several applications utilize mobile sensors.  For example, ZebraNet is
a habitat monitoring application where sensors are attached to zebras
and collect information about their behavior and migration
patterns~\cite{zebranet}.  In addition, applications where sensors are
deployed on humans (e.g., in cellular phones to measure reception
quality and help assess coverage) or vehicles have been suggested. 

A simple algorithm for localization is to do so at a fixed frequency
(for example, this is the algorithm used in the ZebraNet habitat
monitoring application~\cite{zebranet}).  However, using a fixed
frequency may be insufficient if the sensor is moving faster than the
localization frequency can keep track of.  Conversely, if the sensor
is not moving fast, the localization frequency may be overly
aggressive, leading to expensive unecessary localization operations.
To address these effects we propose two new classes of localization
approaches: (1) Adaptive; and (2) Predictive.  Adaptive localization
dynamically adjusts the localization period based on the recent
observed motion of the sensor, obtained from examining previous
locations.  This approach allows the sensor to reduce its localization
frequency when the sensor is slow, or increase it when it is fast.  In
the second approach, we let the sensors estimate the motion pattern of
the node and project this motion in the future.  If the prediction is
accurate, which occurs when nodes are moving predictably, estimates of
location may be generated without localization, allowing us to further
reduce the localization period.  

We propose algorithms that fit the two classes above and compare them
to static, fixed-period, localization both using simulation and
analysis.  We show that dynamic localization can significantly improve
the energy efficiency of localization without sacrificing accuracy in
the location estimation (improving accuracy in most situations).

The remainder of this paper is organized as
follows. 
Section~\ref{related} overviews some related
work.  In Section~\ref{problem} we define the dynamic localization
problem and present candidate protocols for addressing it in
Section~\ref{protocols}.  Section~\ref{analysis} presents some
analysis of the performance of the protocols under special conditions.
In Section~\ref{experiment} we carry out an evaluation study of the
protocols.  Finally, in Section~\ref{conclude} we present some
concluding remarks.

\section{Related Work}~\label{related}

Localization has received a lot of attention in the context of static
sensor networks. The protocols presented in this paper are independent
of the actual localization technique. However, we now mention some of
the state-of-the art techniques which can be used for localization.
He et. al~\cite{he03rangefree} have classified existing localization
techniques into two categories: \textit{range-based} and
\textit{range-free}.In range-based techniques, information such as
distances (or angles) of a receiver are computed for a number of
references points using one of the following signal strength or timing
based techniques and then position of the receiver is computed using
some multilateration technique~\cite{ward97new}. However, range-free
techniques do not depend upon presence of any such information.

Localization techniques typically require some form of communication
between reference points (nodes with known coordinates) and the
receiver (node that needs to localize). Some examples of communication
technologies are RF-based and acoustic based communication. In RADAR
system~\cite{bahl00radar}, RF-based localization is suggested, where
distance is estimated based on received signal
strength. Cricket~\cite{priyantha00cricket} uses concurrent radio and
ultrasonic sounds to estimate distance.  Some researchers have used
Time based techniques such as Time-of-Flight(TOA) ~\cite{ward97new},
Time-Difference-of-Arrival(TDOA)~\cite{priyantha00cricket,ahlos}
between reference point and the receiver node as a way to estimate
distance.  Niculescu et. al~\cite{niculescu03ad} proposed using
angle-of-arrival to estimate position.  Recently He
et. al~\cite{he03rangefree} proposed range-free techniques for
localization.

A straightforward localization approach would make use of Global
Positioning System (GPS). Existing research projects such as
zebra-net~\cite{zebranet} uses a GPS based localization, where mobile
sensors find out their location every three minutes. He
et. al~\cite{he03rangefree} pointed out, GPS based systems require
specialized hardware for precise synchronization
with the satellite's clock. GPS uses one-way flight time information
whereas other systems such as Local Positioning System
(LPS)~\cite{ward97new} use round-trip-time to avoid time
synchronization.

Bulusu et. al~\cite{Bulusu00a} studied signal strength based and
connectivity based techniques for localization in outdoor
environments.  Recently Kumar et. al~\cite{kumar03} proposed using
dead reckoning-Based Location services for mobile ad-hoc
networks. However, to the best of our knowledge this paper is the
first attempt to apply such predictive techniques for localization in
mobile sensor networks.

\section{Problem Definition}~\label{problem}

At every {\em
  localization point}, the node invokes its localization mechanism
  (e.g., using GPS, triangulation based localization, or otherwise) to
  discover its current location $(x_i,y_i)$.  The {\em localization
  point vector} is the sequence of localization points collected by a
  sensor is denoted $S_{i}$.  We assume that the localization
  mechanism estimates the current position with a reasonable
  tolerance.  In the figure, the uncertainty introduce by the
  localization mechanism is represented by the small circles.  

In the time duration between two consecutive localization points, the
error in the estimate of the location increases as the node moves (on
average) increasingly further from its last location estimate.  In
order to control this error, localiztion must be repeated with enough
frequency to ensure that the location estimate meets some
application-level error requirements (e.g., the estimate remains
within a prespecified threshold from the actual location).  However,
carrying out localization with high frequency drains the node's
energy.  Solutions to this problem must balance the need to bound
error with the cost of carrying out localization.  Exploring protocols
that effectively estimate location while minimizing the localization
operations is the problem we consider in this paper.

  We keep our analysis independent of the specific localization
  mechanism used.  Note that dynamic control of localization is needed
  whether localization is carried out on demand (i.e,, the node
  queries neighbors or fixed localization nodes for localization
  information) or proactively (e.g., by having localization nodes
  periodically transmit localization beacons, or using GPS).  If
  localization is on-demand , the localization mechanism can be
  invoked when needed.  Alternatively, if the localization is done
  periodically without control of the sensor node, the node can still
  control its localization frequency by deciding when to start
  listening to the beacons.  Since receiving packets or GPS signals
  consumes significant energy, controlling the localization frequency
  also applies for such schemes.

The primary tradeoff is between the observed localization error and
the energy consumed. The localization error stands for diveregence of
reported location from actual location. We measure divergence in terms
of euclidean distance between actual and reported coordinates -- we
term this the {\em absolute error}. We also consider a threshold based
error metric where we compare the absolute error to an appplication
defined tolerance distance ($dist_{tolerance}$); a localization error
lower than tolerance distance is acceptable to the application.  We
measure the percentage of the time that the localization estimate is
within the application defined threshold.

\section{Dynamic Localization Protocols}~\label{protocols}

In this section, we introduce the proposed protocols for dynamic
sensor localization.  We evaluate three approaches for dynamic
localization: (1) Static localization: the localization period is
static; (2) Adaptive localization: the localization period is adjusted
adaptively, perhaps as a function of the observed velocity which can
be approximated using the last two localization points; and (3)
Predictive localization: in this approach, we use dead reckoning to
project the expected motion pattern of the sensor based on the recent
history of its motion.  In the remainder of this section, we introduce
our proposed protocols for each of these approaches in more detail.

\noindent
{\bf Static Fixed Rate (SFR):} This is the base protocol where
 localization is carried out periodically with a fixed time period
 $t$.  This protocol is simple and its energy expenditure is
 independent of mobility; however, its performance varies with the
 mobility of the sensors.  Specifically, if a sensor is moving
 quickly, the error will be high; if it is moving slowly, the error
 will be low, but the energy efficiency will be low.

\noindent
{\bf Dynamic Velocity Monotonic (DVM):} In this adaptive protocol, a
 sensor adapts its localization as a function of its mobility: the
 higher the observed velocity, the faster the node should localize to
 maintain the same level of error.  Thus whenever a node localizes, it
 computes its velocity by dividing the distance it has moved since the
 last localization point by the time that elapsed since the
 localization.  Based on the velocity, the next localization point is
 scheduled at the time when a prespecified distance will be travelled
 if the node continues with the same velocity.  This distance, for
 example, can be the application specified desired maximum error
 threshold.  Thus, when the node is moving fast, localization will be
 carried more often; when it moves slowly, localization will be
 carried out less frequently.

In this protocol, there is a settable parameter $\alpha$ that
represents the target maximum error.  At every localization point, the
current estimated velocity is computed.  Based on this value we
estimate the time that the target maximum error will be reached if the
node continues with the same velocity -- the next localization point
is scheduled at that point.  Note that this approach assumes that a
node is moving with a constant velocity between localization points.
This may not be always accurate -- for example, if a node was standing
still for half the period, then started moving at a velocity $v$, the
estimated velocity will be $\frac{v}{2}$, and we will end up with
suboptimal localization (e.g., exceeding the error threshold for some
time).  Moreover, for very low speeds the localization period may be
computed adaptively to be very large (e.g., a period of infinity would
be predicted if the node is standstill).  Similarly, if the speed is
very high, the localization period may become very low, wasting a lot
of energy.  To account for these effects, we place an upper and a
lower limit on the localization periods.  The effect of these is
explored in the analysis section.

\noindent
{\bf Mobility Aware Dead Reckoning Driven (MADRD):} This is a
 predictive protocol that computes the mobility pattern of the sensor
 and uses it to predict future mobility.  Depending on how well the
 mobility of the sensor can be predicted, the localization frequency
 can be signficantly reduced using this approach.  To the best of our
 knowledge, this is the first paper to appy dead reckoning for
 localization in mobile sensor network.

Using dead reckoning localization should be triggered when the
expected difference between the actual mobility and the predicted
mobility reaches the error threshold.  This is in contrast to DVM
where localization must be carried out when the distance from the last
localization point is predicted to exceed the error threshold.  Thus,
if the node is moving predictably, regardless of its velocity,
localization can be carried out at low frequency; if the predicted
mobility pattern is perfect and holds for all future time, no further
localization would be necessary. 

\subsection{Predicted Mobility Pattern}

The predicted mobility pattern will generally be imperfect due to the
following reasons: the developed model can be inaccurate -- the
sampled points may not be sufficient to discover the mobility pattern.
Furthermore, we may assume an inapporpriate mobility model (e.g.,
assuming that the node is moving at constant velocity when it has an
acceleration component). In addition, since the localization mechanism
introduces some error in the computed localization points, even if we
have sufficient samples and the assumed model matches the true
mobility pattern we will end up estimating mobility inaccurately due
to the error in the localization points.  Finally, sensors will
typically not follow a predictable model -- for example, there may be
unpredictable changes of directions or pauses that will cause the
predicted model to go wrong.  For all these reasons it is necessary to
continue localization periodically to detect deviations from the
predicted model.  If dead reckoning is carried out aggressively, then
a change in the mobility pattern (for example, a standstill node
starting to move) can cause large errors as the node continues to
predict location based on past behavior.

Thus, there are a number of different protocols that can be
constructed with these properties.  Specifically, these protocols may
differ in how they construct the predicted mobility pattern (e.g., a
first order model that assumes constant velocity between points, or a
second order model that assumes a velocity and acceleration
components).  Moreover, they may differ in how often they localize to
detect variations between the predicted and actual mobility patterns.
We do not explore the full range of such protocols.  Instead, we
select a simple instance of dead reckoning protocols that works as follows.
 
\begin{figure}
\centerline{\epsfig{file=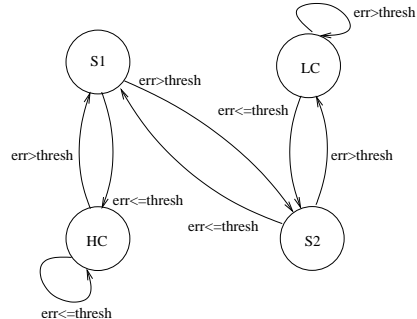,scale=0.5}}
     \caption{State Diagram for Dead Reckoning}\label{state-diagram}
\end{figure}

Accounting for differences between the predicted model and the actual
mobility of the sensor, including errors due to changes in the mobility
pattern that occur after or during dead-reckoning estimation is almost
impossible.  In practice, we use the following approach.  Like DVM,
and for similar reasons, we define maximum and minimum localization
periods.  Moreover, we score the performance of our prediction at
every localization point by comparing the predicted location to the
actual location.  If the prediction is erroneous (larger than a
prespecified rate of divergence), we move towards a low confidence
state and become more aggressive in localization.  The intuition is
that the mobility pattern is changing, and more localization is needed
to capture the new mobility pattern as well as to bound the
localization error.  However, if the prediction is accurate, our
confidence in the predictor increases and we increase the localization
period.

A state diagram for MADRD is shown in Figure~\ref{state-diagram}. In
this diagram, HC refers to the high confidence state where the
predictor is scoring well and localization period is increased.  LC
refers to the low confidence state where the predictor is not scoring
well and the period is decreased.  Erroneous predictions move the
predictor towards the LC, while correct predictions move it towards
HC.  States S1 and S2 provide some hysterisis between LC and HC.

\section{Analysis for Special Cases}~\label{analysis}
In this section, we evaluate SFR and MADRD under the following special
conditions: (1) constant velocity with a turn; and (2) constant
velocity with periods of no motion.

\subsection{Change of Direction Scenario}

\begin{figure}[htb]
	\centering
		\includegraphics[scale=0.35]{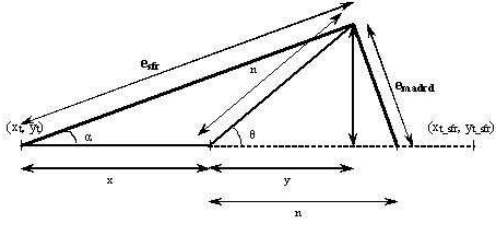}
	\caption{Error if deviation of $\theta$ degrees is taken}
	\label{fig:stLineDev}
\end{figure}

\begin{figure}[htb]
	\centering
		\includegraphics[scale=0.35]{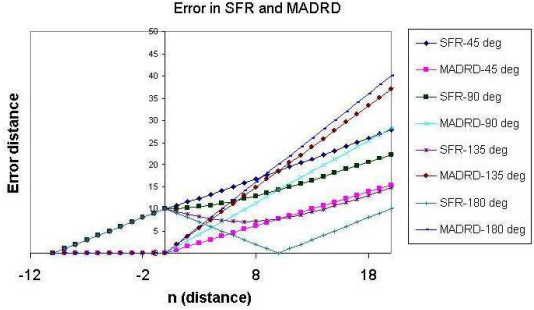}
	\caption{Errors in SFR and MADRD}
	\label{fig:errAnalysis}
\end{figure}

Now consider the node taking the deviation of $\theta$ degrees. Let
the distance at which the node takes the deviation be $d$ meters after
the localization point $(x_t,y_t)$. The time at which the deviation
occurs is greater than time $t$ and lesser than $t+t_{sfr}$. Figure
\ref{fig:stLineDev} shows the movement of the node. The distance $x+y$
signifies the distance covered in time $t_{sfr}$ with constant
velocity $v$.

The error in localization between time $t$ and $t+t_{sfr}$ can be
split up into two parts. The first part is error before the deviation
occurs (identical to the fixed velocity analysis above) and the second
one is after the deviation. Let $n$ be any point on the expected line
of motion that the node would have travelled if it had not taken the
deviation. If the node would have travelled a distance of $n$ along
expected straight line, it will travel the same distance after
deflection because of constant velocity. Let $n=0$ at the point of
deviation and increases along the straight line.

\begin{figure*}[ht]
\begin{center}
\subfigure[Speed (4-5 m/s)(Upper Threshold 6 sec)\label{madrdmove}]{\epsfig{file=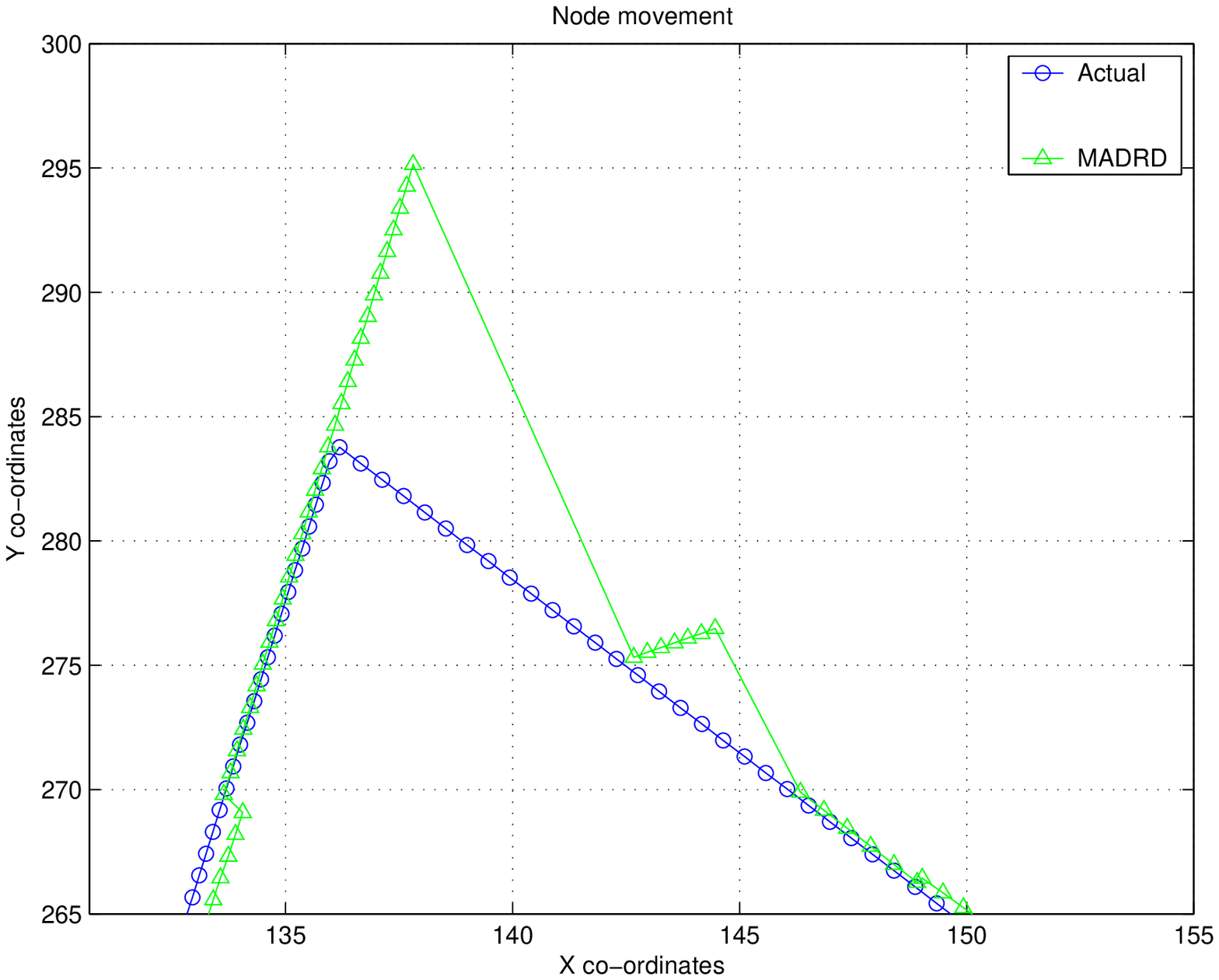,scale=0.4}}
\subfigure[Speed (0.5-1 m/s)(Upper Threshold 10 sec)\label{madrdpause}]{\epsfig{file=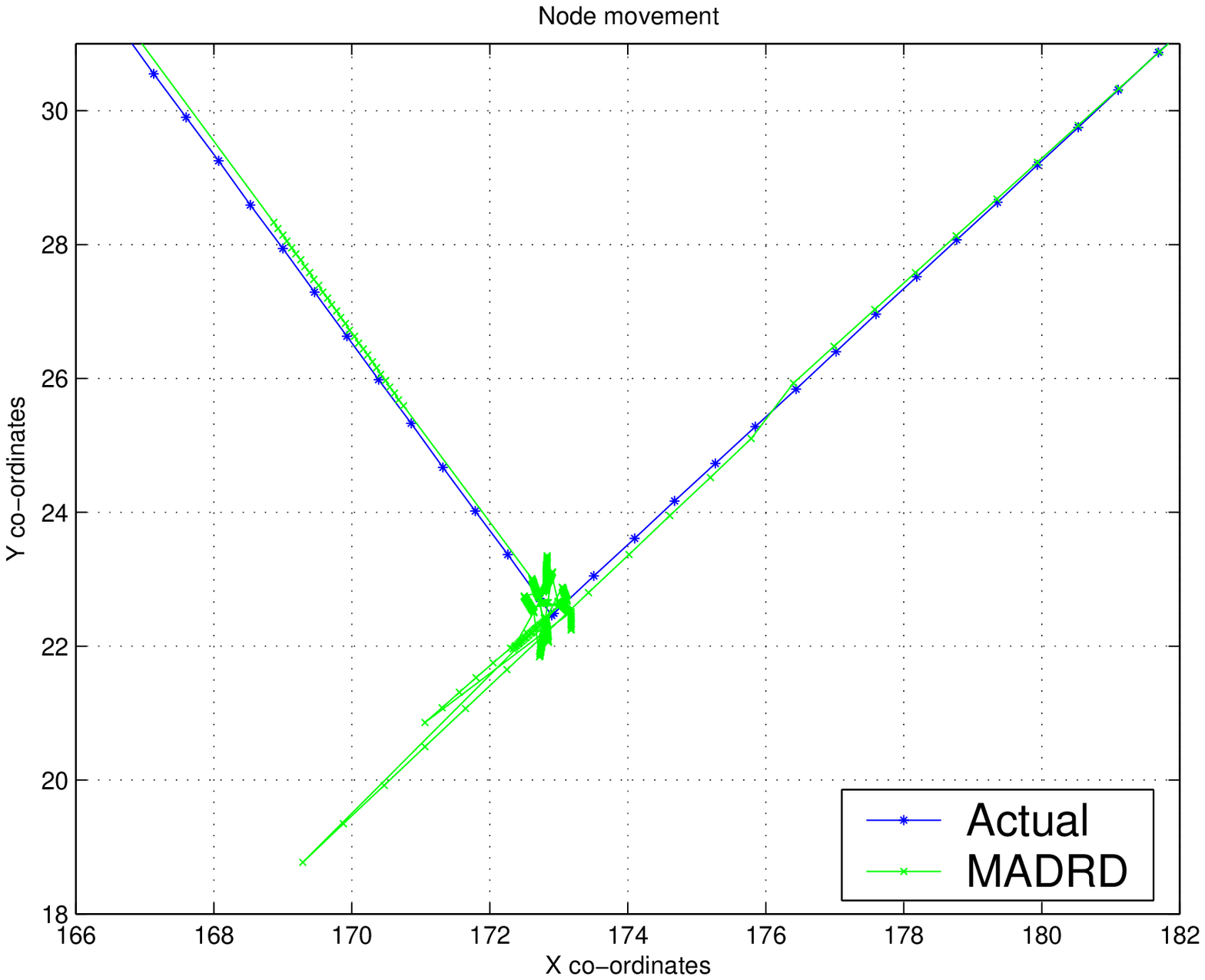,scale=0.4,silent=}} \quad

\caption{MADRD Behavior with Unexpected Change in Velocity}
\label{rwayeng} 
\end{center}
\end{figure*}         

\subsubsection{SFR protocol}

Let the node use SFR protocol for localizing. The the error at point
$n$ will be the length of line $e_{sfr}$ shown in Figure
\ref{fig:stLineDev}. The equation for $e_sfr$ is given by Equation
\ref{eq:e_sfr}.

\begin{equation}
\label{eq:tanalpha}
\tan{\alpha} = \frac{n\times\sin{\theta}}{(x+n\times\cos{\theta})}\\
\end{equation}

\begin{equation}
\label{eq:e_sfr}
e_{sfr} = \frac{n\times\sin{\theta}}{\sin{\alpha}}
\end{equation}

Figure \ref{fig:errAnalysis} shows the graph of error against $n$ . As
$n$ increases from $0$ to $y$, the $e_{sfr}$ varies as shown in the
graph in Figure \ref{fig:errAnalysis}. We can see that for $n>0$, the
curve is not linear. This can be seen clearly in the case where
$\theta=135$.

As the angle of deflection increases from $0\ degrees$ to $90\
degrees$, the error in SFR decreases because the line of motion will
be nearer to $(x_t,y_t)$ when $\theta$ increases. For angles greater
than $90\ degrees$ and lesser than $180\ degrees$. The error decreases
as node moves towards $(x_t,y_t)$ and then starts increasing.

At $\theta=180\ degrees$, the error touches zero after the node has
covered $x$ distance and then the error starts increasing
linearly. Now the error vector is in other direction than the earlier
error vector. Graph in Figure \ref{fig:errAnalysis} shows the absolute
value of the error.

\subsubsection{MADRD protocol}
\begin{equation}
\label{eq:e_madrd}
e_{madrd} = {2 \times n \times \sin{\frac{\theta}{2}}}
\end{equation}
The length of the line $e_{madrd}$ in Figure \ref{fig:stLineDev} shows
the the error in MADRD protocol. It increases linearly as the $n$
increases. This is given by the equation \ref{eq:e_madrd}. Graph in
Figure \ref{fig:errAnalysis} shows the comparison of MADRD protocol
with SFR for different angles. We observe for acute angles, MADRD
protocol performs better than the SFR. However, if $\theta$ is between
$90\ degrees$ and $270\ degrees$, SFR starts performing better. This
is because the node is moving away from the predicted motion line and
$e_{sfr}$ is smaller than the $e_{madrd}$.

\subsection{Pause Scenario}

In this case, the node comes to a standstill after being in motion
with velocity $v$.  Let the distance at which the node stops be $d$
meters after the localization point $(x_t,y_t)$, but before the next
localization point.  In this case, the error in SFR increases linearly
until $d$, when it stops increasing.  Conversely, the error in MADRD
starts at 0 while the node maintains the speed of $v$.  However, when
it stops moving, the error in MADRD starts increasing proprtionately
to $v$ since the predictor assumes that the node continues in motion.
Interestingly, if the node is standstill but suddenly starts moving
with velocity $v$, SFR and MADRD will behave identically until the
next localization point (which may be different for each).  The reason
is that SFR's uses the implicit prediction that the node remains at
the point of the last localization.  In this scenario, MADRD uses the
same predictor since the node actually was not moving at the last
localization point.

Figure~\ref{madrdmove} shows the behavior of MADRD when a turn occurs.
In this case, the MADRD estimate continues predicting motion in the
original direction.  Moreover, even when localization occurs, the
average velocity computed as a predictor for the next period will be
off as well (it represents the weighted average of the original as
well as the new velocities).  A similar trend is observed in
Figure~\ref{madrdpause} where a node pauses after being in motion at a
constant velocity.  In this case, the MADRD estimate overshoots the
node along the old trajectory when it pauses.

\section{Experimental Results}~\label{experiment}
In this section we present the results of our experiments with the
proposed protocols.  In order to analyze the protocols, we use the
ns-2 discrete event simulator~\cite{ns-2}.  We use a simulation area
of 300 by 300 meters, with sensor transmission range of 100 meters
using IEEE 802.11.  We use 36 equally spaced beacon nodes for
localization and 24 mobile nodes carrying out localization.  Each
simulation was run for 900 seconds.    We
use a query based localization mechanism: a node that is interested in
localization broadcasts a request -- beacons that receive the request
reply with their location which can then be used to triangulate the
nodes own location.  The beacons are placed such that at least three,
and sometimes four, beacons are able to answer each query.  Please
note that our results are not dependent on this localization model: we
measure the energy in terms of number of localization operations,
regardless of how the localization is carried out.

\begin{figure}
\begin{center}
\epsfig{file=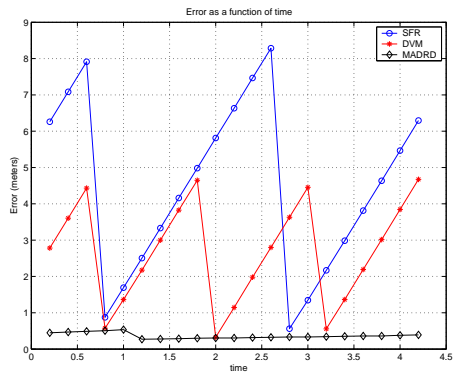,scale=0.35}
\caption{Instantenous Error for Speed (4-5 m/s).}
\label{errn0} 
\end{center}
\end{figure}                        

\begin{figure*}
\begin{center}
\subfigure[Speed (0.5-1 m/s)(Upper Threshold 10 sec)\label{fig1}]{\epsfig{file=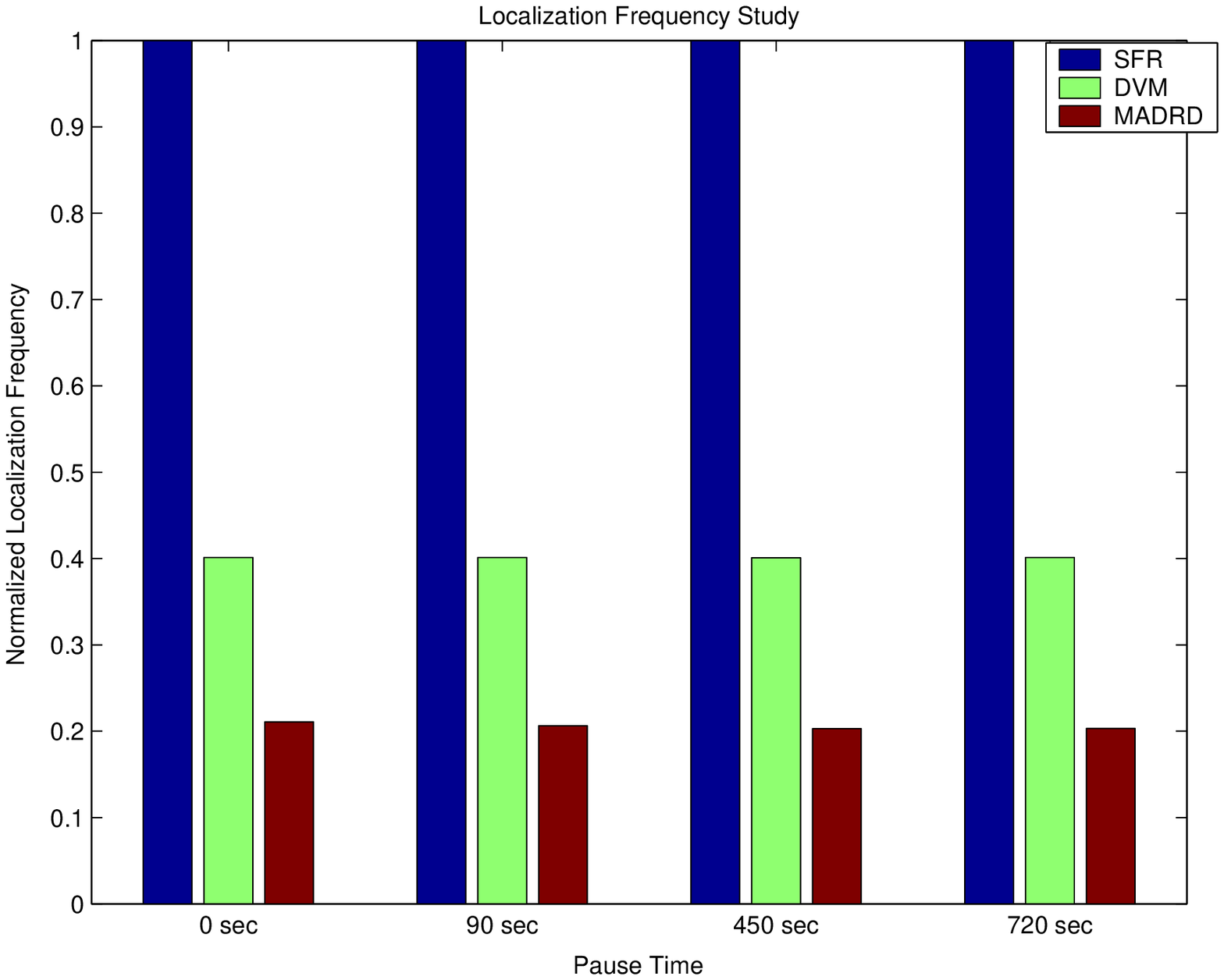,scale=0.35,silent=}} 
\subfigure[Speed (4-5 m/s)(Upper Threshold 6 sec)\label{fig2}]{\epsfig{file=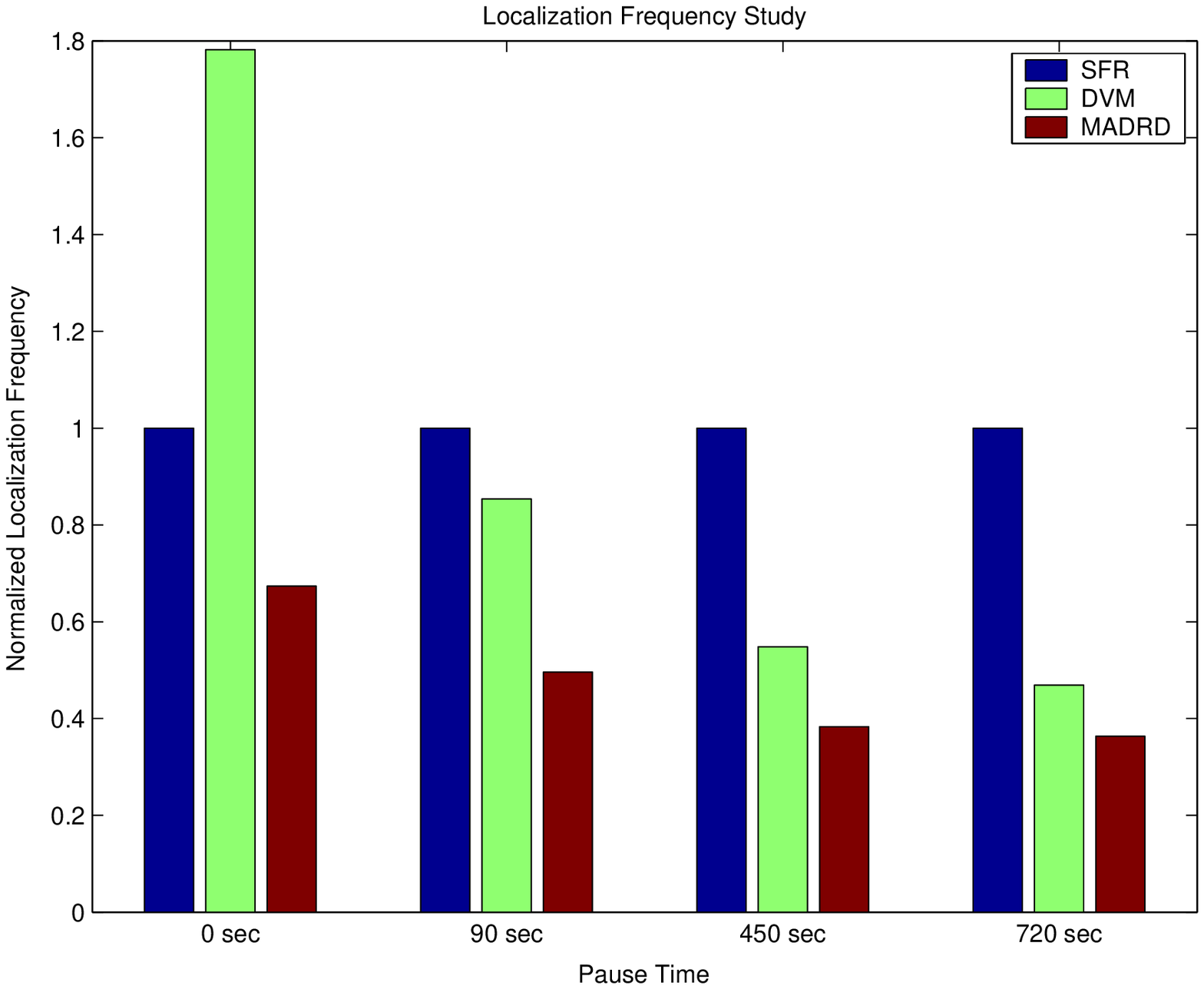,scale=0.35}}
\end{center}
\caption{Localization Frequency as a function of  mobility and pause time.}
\end{figure*}         

\begin{figure}
\begin{center}
\epsfig{file=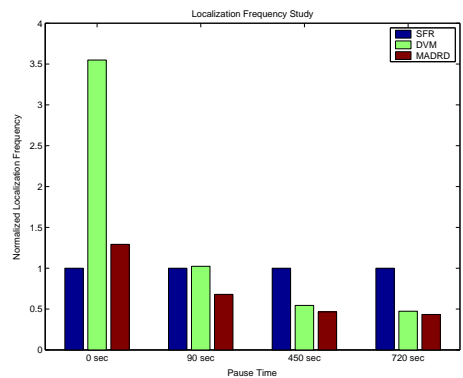,scale=0.35}
\caption{Localization Frequency: (8-10 m/s)}\label{fig3}
\end{center}
\end{figure}

\begin{figure}
\begin{center}
\epsfig{file=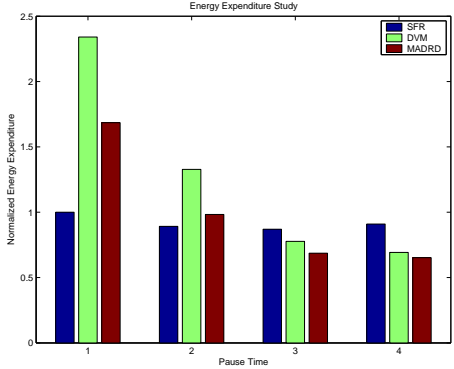,scale=0.35}
\caption{Localization Energy: (8-10 m/s)}\label{fig4}
\end{center}
\end{figure}

The assumed mobility model has significant implications on the
performance of the localization protocols. 

We consider first the random waypoint model, widely used in the mobile
ad hoc network community.  In this model, a node picks a random
location in the simulated area and starts moving to it with a
controllable average velocity.  When the node reaches the destination,
it pauses for some fixed pause time.  The model is predictable while
the node is moving, or for the duration of the pause but not during
the period where it pauses or when it starts moving.  Further, if the
pause time is zero, the model is unpredictable when the node reaches
its destination, then picks another randomly and starts moving towards
it.  We can control how predictable the model is by manipulating the
average speed and the pause time -- if the pause times are short, the
node has more unpredictable behavior.  Finally, we present some
limited results with Gaussian Markovian mobility pattern which does
not lend itself well to prediction using a constant velocity model as
we do in MADRD. We used BonnMotion tool~\cite{bm} to generate the 
various scenarios.

Figure~\ref{errn0} shows the Instantaneous error for random way-point
mobility model with speed uniformly distributed between 4-5 m/sec.
The SFR period in this case was chosen to be 2 seconds -- the node
invokes localization once every two seconds.  Note that in the case of
SFR and DVM the node assumes that the last measured localization point
is its current location. Therefore $Error_{Inst_{t}}$ continues to
grow between two successive localization points as the node moves away
from its last localization point.  Figure \ref{errn0} shows the
instantenous error for SFR, DVM and MADRD protocols. In the case of
SFR, sensor 0 localizes approximately at times 0.6, 2.6.  As one can
see upon localization the error lies within the localization mechanism
error range (which we picked to be uniformly distributed between 0 to
0.5 meters). In between the two localization points, the error
increases linearly up to 8 meters. In the case of DVM, a similar trend
is seen again, howerver due to adaptive localization intervals, the
magnitude of the error is lower than that of SFR; DVM was able to
discover that it needs to localize more often than once every 2
seconds. In the case of MADRD protocol, the ability to predict the
current location gives rise to very low error since the node actually
follows the prediction.  This graph clearly shows the strength of
dead-reckoning procotols due to their prediction capability.

\begin{figure*}
\begin{center}
\subfigure[Error Comparison\label{abserr}]{\epsfig{file=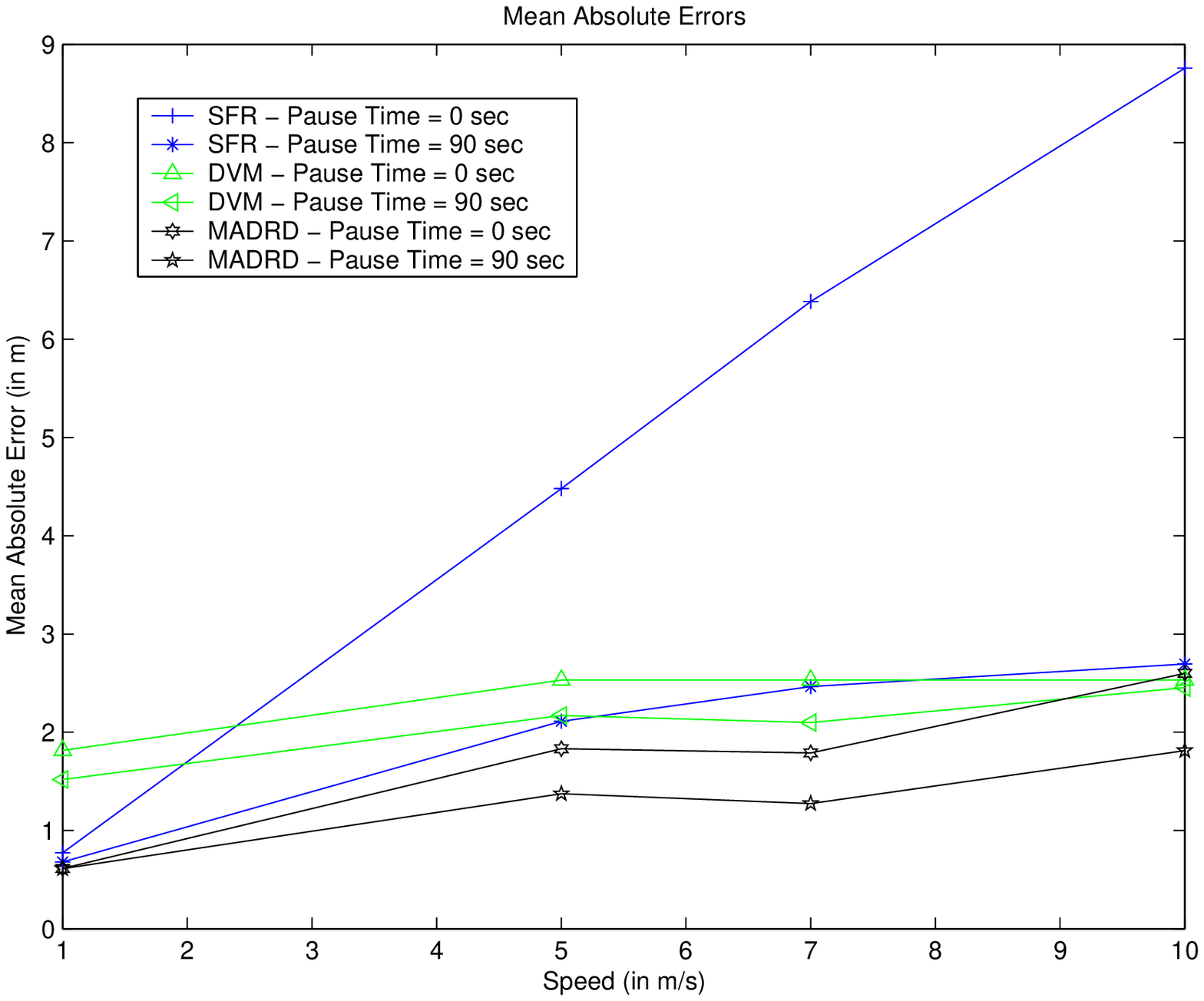,scale=0.35,silent=}} \quad
\subfigure[Error vs. Pause Time\label{abserrbar}]{\epsfig{file=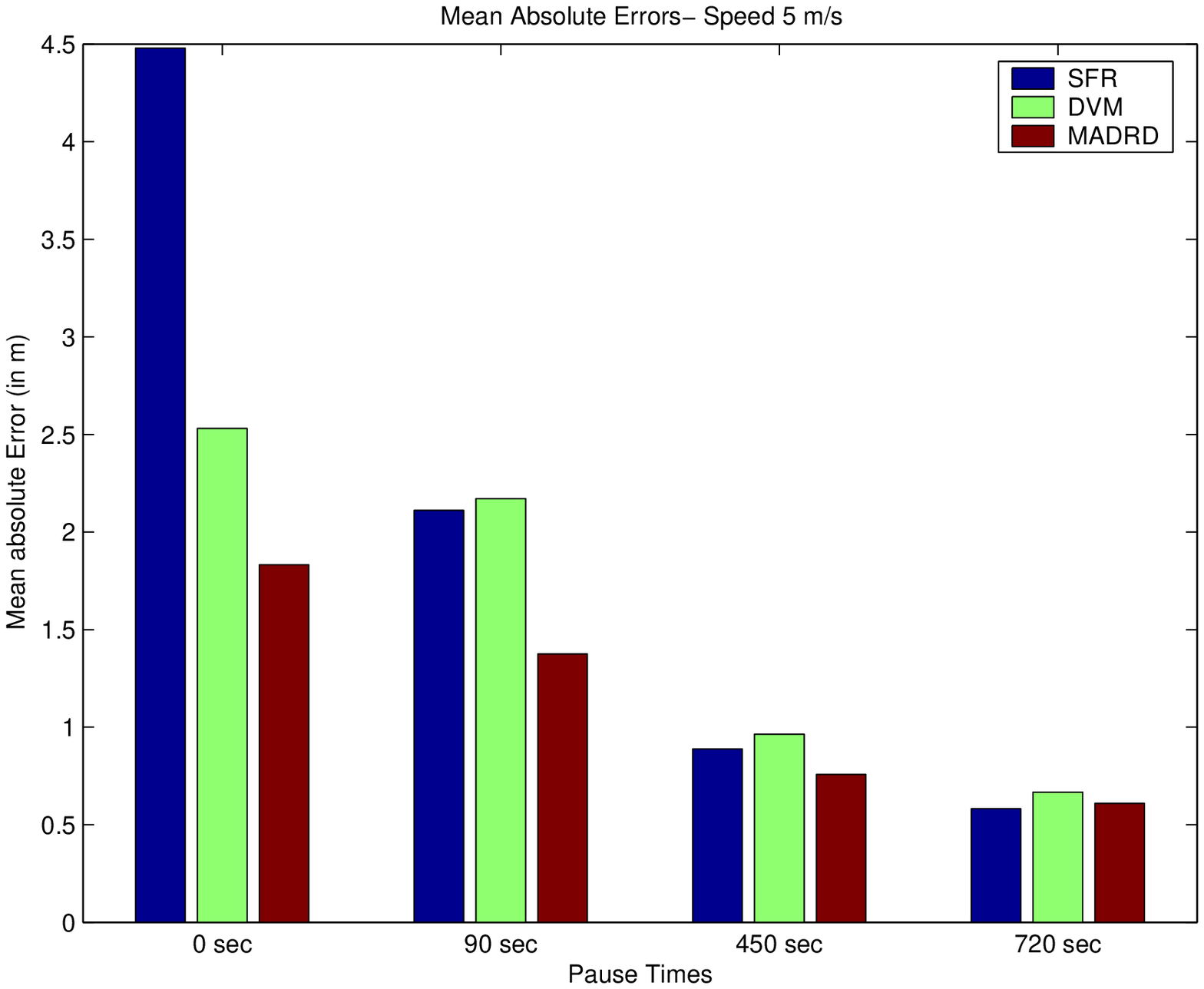,scale=0.35}}
\label{Absolute Error as a function of Mobility and Pause Time}
\end{center}
\end{figure*}
 
Figures~\ref{fig1}, \ref{fig2} and \ref{fig3} show the number of
localization operations for three
different average velocities normalized to the number needed by SFR.
The number of localization operations correlates directly with
localization energy since the average cost of localization is constant
for most localization schemes.  This fact is highlighted in
Figure~\ref{fig4} which shows the energy expenditure for the same
scenarios as in Figure~\ref{fig3} -- the shapes of the figures are
very similar.  In the case of low mobility \ref{fig1}, DVM and MADRD
localize less often than SFR.  However, as the speed increases, the
energy expeniture of DVM and MADRD grow more than that of SFR. Note
that since these protocols are adaptive, even for high speeds they
adapt well with the increase in pause time thereby spending less
energy than SFR when pause time is high.

\begin{figure*}[ht]
\begin{center}
\subfigure[Accuracy Comparison\label{absacc}]{\epsfig{file=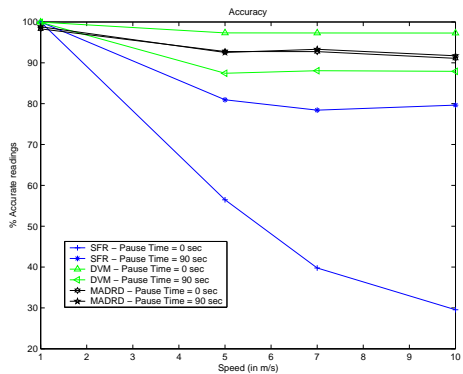,scale=0.35,silent=}} \quad
\subfigure[Accuracy vs. Pause Time\label{absaccbar}]{\epsfig{file=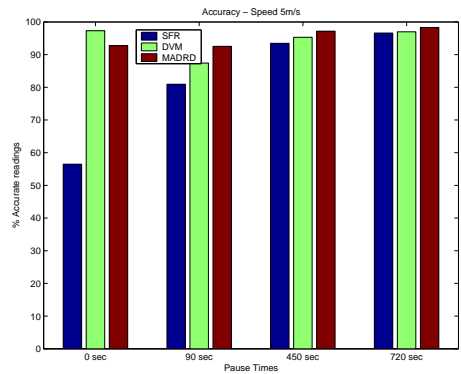,scale=0.35}}
 
\caption{Percentage accuracy study a function of  mobility and pause time.}
\end{center}
\end{figure*}         

Figure~\ref{abserr} shows the absolute error as a function of mobility
for the four protocols for two different pause time values.  The
primary observation here is that the error for SFR grows linearly with
the average velocity while both DVM and MADRD manage to adapt their
localization and maintain an error that does not grow significantly
with the velocity. Note that under high mobility, this requires more
localization operations than SFR as was reflected in the localization
frequency diagram for high speeds and low pause time.
Figure~\ref{abserrbar} shows the effect of pause time for one
velocity.  Since pauses affect the prediction of DVM and MADRD, their
advantage in terms of error relative to SFR is highest with no pause
time.  At very high pause times, all three protocols perform well.
An alternative measure of localization effectiveness is to monitor the
fraction of the simulation time where the localization estimate was
within an application specified threshold (in this case 5 meters).
Figure~\ref{absacc} shows the accuracy as a function of mobility for
two pause times.  Again, the same trend observed in error is observed
here -- DVM and MADRD perform much better than SFR, especially as
mobility grows.  Figure~\ref{absaccbar} shows the accuracy for one
average velocity as the pause time is varied.

\begin{figure*}[ht]
\begin{center}
\subfigure[MADRD speed (4-5 m/s)\label{uabserr}]{\epsfig{file=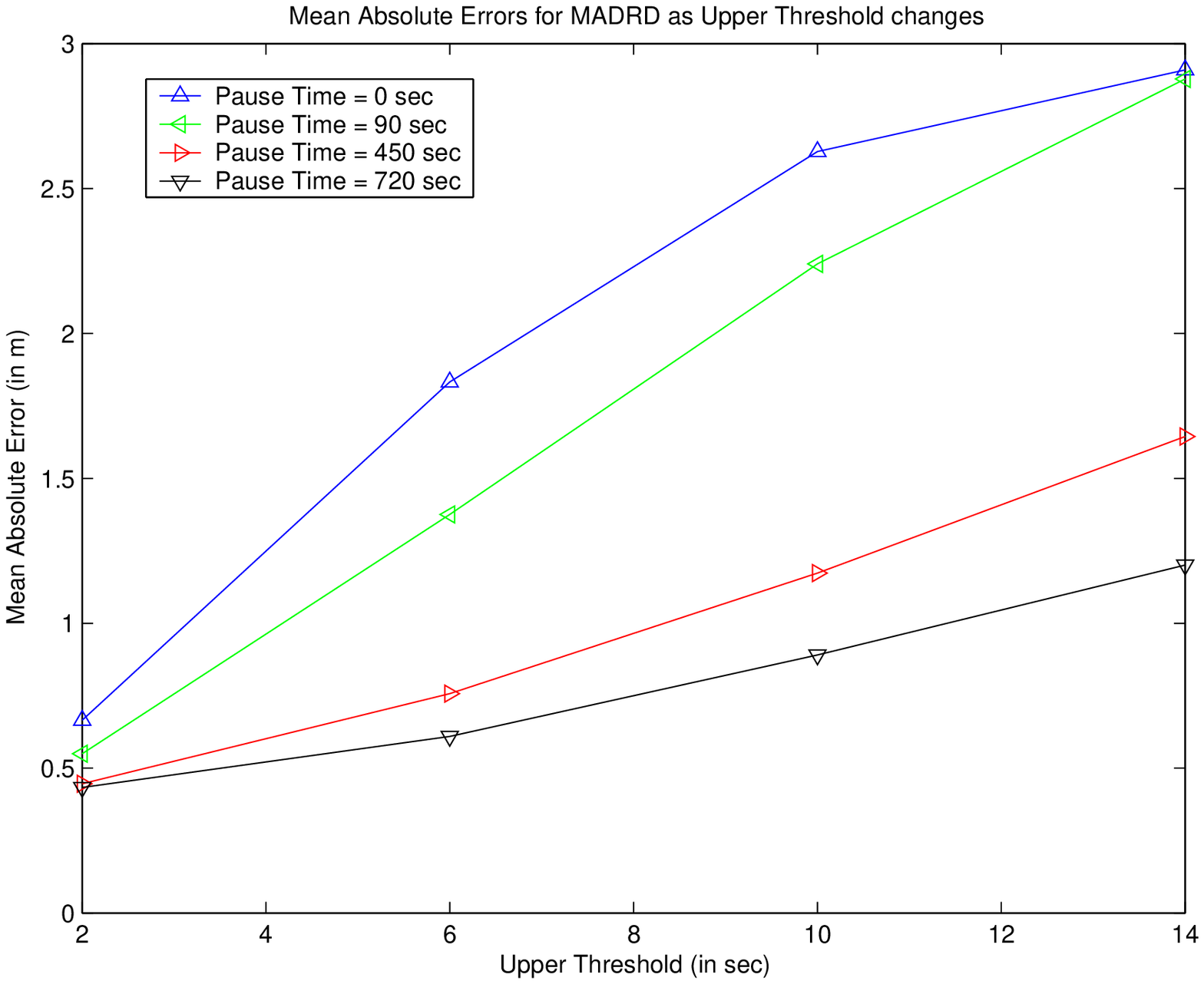,scale=0.35,silent=}}\quad
\subfigure[MADRD speed (4-5 m/s)\label{uabserr2}]{\epsfig{file=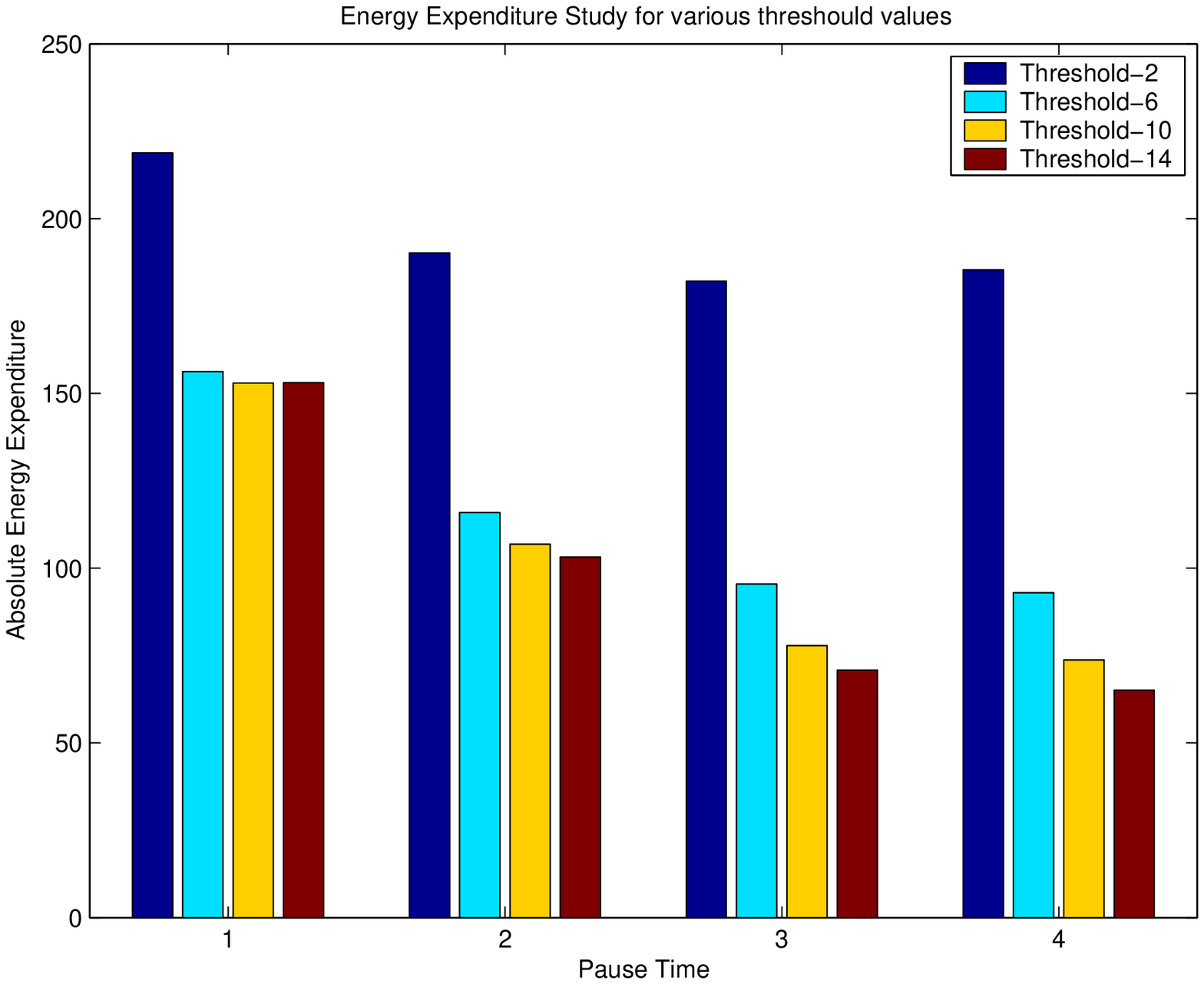,scale=0.35,silent=}}
\caption{Performance of MADRD protocol as a function of Upper Query Threshold for a fixed mobility}
\label{madrduvaruation} 
\end{center}
\end{figure*}

  Recall that to protect against inaccuracies in the prediction model
or unexpected changes in the mobility model MADRD must limit the
maximum period between localizations (upper query threshold).
Figure~\ref{madrduvaruation} shows the effect of this tradeoff -- we
vary the upper query threshold and observe the effect on the accuracy,
error and localization energy.  If the threshold is raised, this
allows MADRD to aggressively predict location without forcing
localization operations to ensure that the predictions are accurate.
Thus, at high thresholds, higher energy savings are possible, but the
expected error grows.  A good value for the upper threshold must
balance these two effects.  Finally, we can use backtracking as
explained in the protocol section to recover from some erroneous
localizaiton estimates.
                                                          
\begin{figure*}[ht]
\begin{center}
\subfigure[Energy: Gaussian Mobility (0.5-1 m/s)(Upper Threshold 6 sec)\label{gausseng}]{\epsfig{file=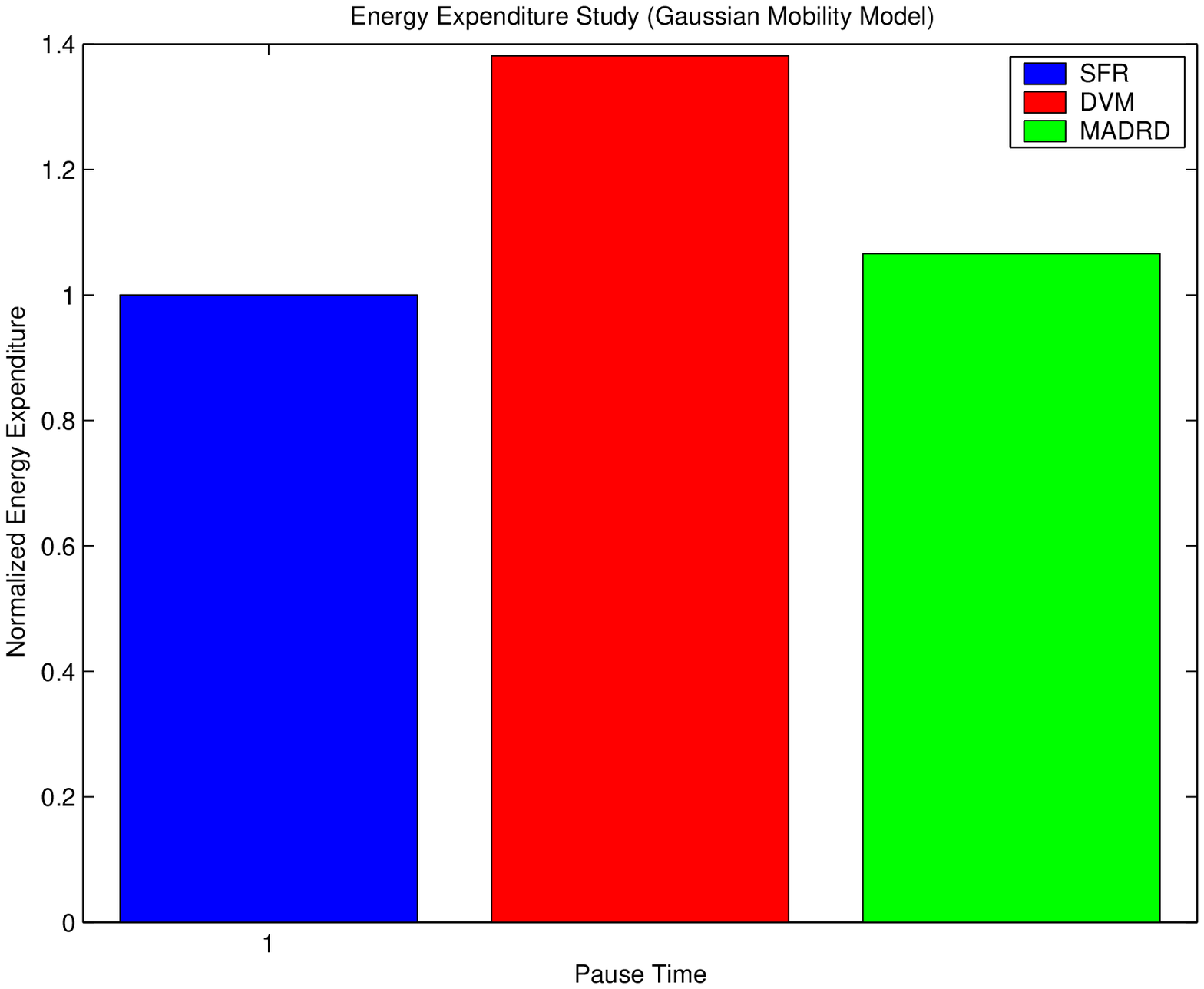,scale=0.35,silent=}} \quad
\subfigure[Error: Gaussian Mobility (4-5 m/s)(Upper Threshold 6 sec)\label{gaussacc}]{\epsfig{file=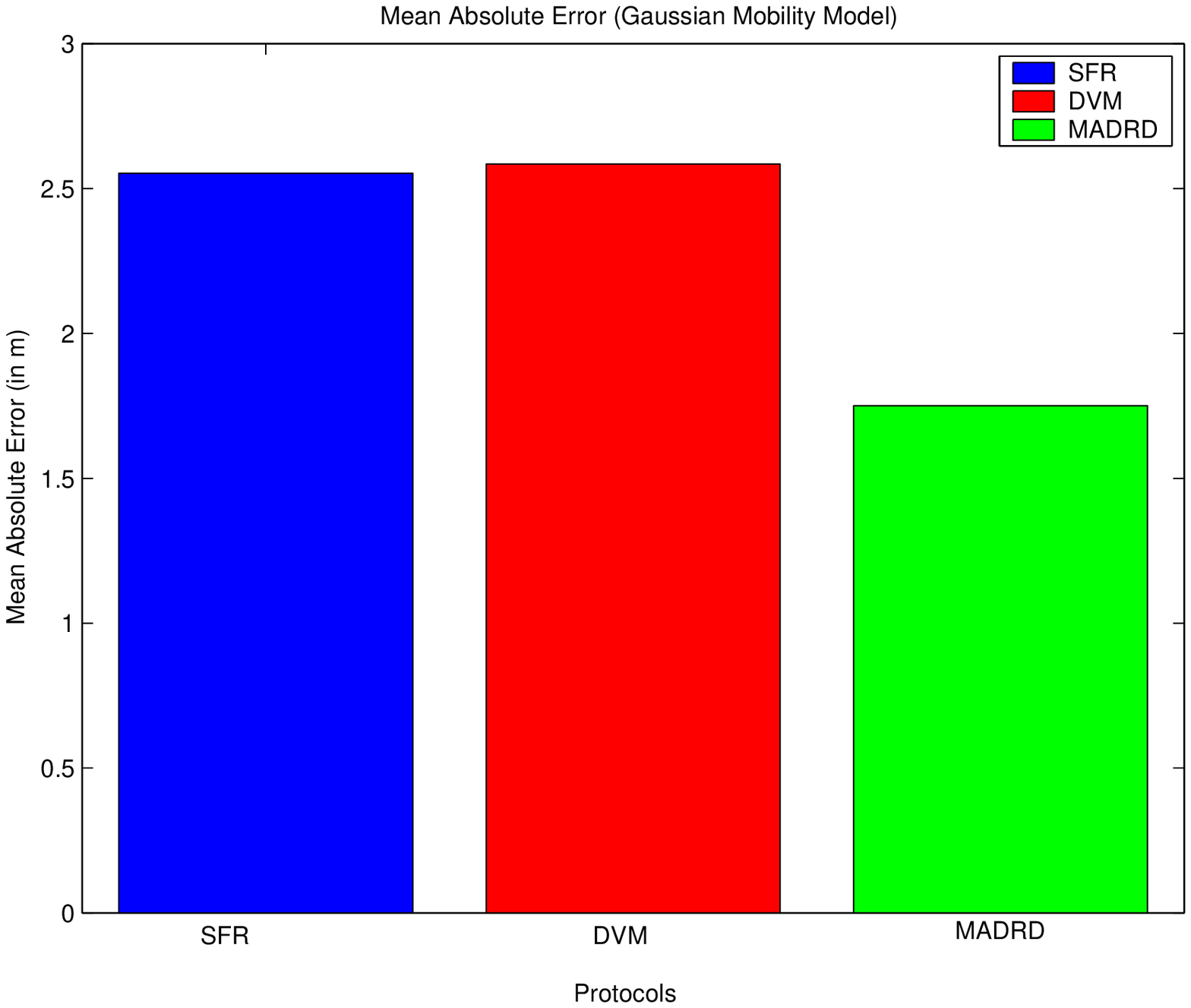,scale=0.35,silent=}}
\caption{Characterizing protocol behavior for Gaussian Mobility Model.}\label{gaussres} 
\end{center}
\end{figure*}

Finally, in Figure~\ref{gausseng} and Figure~\ref{gaussacc} we
evaluate the algorithms using the Gaussian mobility model from an
energy and error perspective.  The Gaussian model is quite different
from the assumed monotonic velocity model that underlies both DVM and
MADRD; thus, this represents one of the worst case scenarios for these
protocols.  Nonetheless, they continue to perform comparably to SFR
(better in most cases), even for this inappropriate mobility model.
In practice, we expect each node to have multiple predictors and
continue to score them.  At each time, the predictor which has
recently been scoring highest would be used to generate the next
localization period.

\section{Concluding Remarks}~\label{conclude}

In this paper, we explored approaches and tradeoffs to the problem of
dynamically managing the Localization period for mobile devices.
Localization has several applications both in Sensor Networks and
Mobile Ad hoc Networks; for example, accurate localization is
necessary to effectively interpret sensor data collected by mobile
sensors.  A basic localization scheme would simply localize
periodically, with a fixed period.  However, since the period is not
sensitive to the actual mobility of the node, the selected period may
be too agressive (wasteful) or insufficient to localize accurately.

We explored two algorithms for dynamic localization: (1) DVM: an
adaptive algorithm that matches the localization period to the
observed velocity of the node; and (2) MADRD: a predictive algorithm
that uses dead reckoning to estimate the location of a node assuming
it is following its recently tracked tracjectory.  We characterized
the performacne of these algorithms for two mobility patterns under
different velocities and pause times.  Both proposed approaches
significantly outperform static localization both from an energy and
accuracy perspectives.  In particular, MADRD performance was excellent
in almost all situations that were studied; however, it is best suited
to mobility patterns that are predictable and this result may not
generalize to other mobility scenarios.

In all three types of protocols, especially the adaptive and
predictive ones, unexpected mobility behavior of the nodes can cause
erroneous localization.  If such situations are to be minimized,
highly aggressive (and inefficient) localization would be needed.
Conversely, if some errors can be tolerated, we can adapt the
localization period more aggressively resulting in significant energy
savings.  We propose a technique called {\em backtracking} to allow
temporary recovery from errors.  Specifically, once localization is
carried out we may discover that the measured location is far from the
expected one.  In this case, it is possible to update the location
estimate after the fact (e.g., using linear interpolation between the
last two points).  This is straightforward for samples that have not
been sent yet.  However, for data that has already been sent, this
requires sending a correction signal.  Since this signal costs energy,
we should still strive to minimize the amount of backtracking needed
by the protocols.  Another future approach to address the same problem
is to use feedback from motion sensors (e.g., an accelerometer).  If
such a device is available, it can be used to interrupt the primary
protocol when a change in the mobility pattern is suspected, causing
it to drop back to training mode to capture the new mobility pattern.

In the future we would like to implement these protocols on existing
sensor prototypes (eg. Motes) and study their performance. The Zebranet
project has developed a simulator for studying systems tradeoffs in
wild-life tracking environment in a realistic setting. We would like
to port our protocols from ns-2 to ZNetSim~\cite{zebranet} and study
the performance for an existing application.  At present our work is
limited to individual mobility models; but in the future we will also
explore group mobility models. For military scenarios for example we
can imagine a group of soldiers moving together to achieve certain
goal. We would like to evaluate the protocols proposed in this paper
for such scenarios and suggest some improvements.

\bibliography{localization,ieee,ref}
\bibliographystyle{acm}
\end{document}